\documentclass{IEEEtran}
\usepackage{mathtools}
\usepackage{amsmath}
\usepackage{blindtext}
\usepackage{bbm}
\usepackage{comment}
\usepackage{caption}
\usepackage{subcaption}
\usepackage{amsmath,xcolor,amssymb,multirow,cite,epsfig,graphicx,graphics}
\usepackage{algorithm}
\usepackage{algorithmic}
\usepackage{makecell}
\usepackage{color, colortbl}

\newcommand{\ar}[1]{\textcolor{black}{#1}}   
\newcommand{\myblue}[1]{\textcolor{black}{#1}}   

\providecommand{\keywords}[1]
{
  \small	
  \textbf{\textit{Keywords---}} #1
}

\newcommand{\specialcell}[2][c]{%
  \begin{tabular}[#1]{@{}c@{}}#2\end{tabular}}

\renewcommand{\baselinestretch}{1.08}

\usepackage[T1]{fontenc}
\usepackage[utf8]{inputenc}
\usepackage{authblk}

\begin{document}



\title{\LARGE Fully-echoed Q-routing with Simulated Annealing Inference for Flying Adhoc Networks} 


\author[1]{Arnau Rovira-Sugranes}
\author[1]{Fatemeh Afghah}
\author[2]{Junsuo Qu}
\author[1]{Abolfazl Razi}
\affil[1]{School of Informatics, Computing and Cyber Systems, Northern Arizona University, Flagstaff, USA}
\affil[2]{School of Automation, Xi’an University of Posts and Telecommunications, Xi’an, China}
\affil[1]{\textit {\{ArnauRovira, Fatemeh.Afghah, Abolfazl.Razi\}@nau.edu}}
\affil[2]{\textit {qujunsuo@xupt.edu.cn}}
\renewcommand\Authands{ and }

\maketitle

\begin{abstract}
Current networking protocols deem inefficient in accommodating the two key challenges of \ar{Unmanned Aerial Vehicle} (UAV) networks, namely the network connectivity loss and energy limitations. One approach to solve these issues is using learning-based routing protocols to make close-to-optimal local decisions by the network nodes, and Q-routing is a bold example of such protocols. However, the performance of the current implementations of Q-routing algorithms is not yet satisfactory, mainly due to the lack of adaptability to continued topology changes.
In this paper, we propose a full-echo Q-routing algorithm with a self-adaptive learning rate that utilizes Simulated Annealing (SA) optimization to control the exploration rate of the algorithm through the temperature decline rate, which in turn is regulated by the experienced variation rate of the Q-values. Our results show that our method adapts to the network dynamicity without the need for manual re-initialization at transition points (abrupt network topology changes). Our method exhibits a reduction in the energy consumption \ar{ranging from 7\% up to 82\%, as well as a 2.6 fold gain 
in successful packet delivery rate}, compared to the state of the art Q-routing protocols \footnote{This material was based upon the work supported by the National Science Foundation under Grants No. 1755984 and 2008784.}.

\keywords{\myblue{UAV networks, learning-based routing, Q-routing, adaptive networking, energy efficiency.}}

\end{abstract}

\section{Introduction}   \label{sec:introduction}

Flying Adhoc Networks (FANETs), \ar{especially those} composed of Unmanned Aerial Vehicles (UAVs), are becoming increasingly popular in many sensing, monitoring, and actuation applications due to their key features such as free mobility, faster speeds, less human hazards in harsh and risky environments, autonomous operation, larger coverage areas, lower costs, and flexible imaging capabilities. The range of applications is countless and includes but not limited to transportation \cite{transportation}, traffic control \cite{traffic-monitoring}, fire monitoring \cite{huang2020wildfire,shamsoshoara2020aerial}, human action recognition \cite{peng2020fully}, surveillance \cite{surveillance}, border patrolling \cite{border-patrolling}, search and rescue \cite{search-rescue}, disaster management \cite{erdelj2017help}, wireless network connectivity \cite{connectivity}, smart agriculture, and forestry \cite{forestry}. However, many communication and control protocols, \ar{which are} primarily developed for ground networks with somewhat stationary infrastructures deem inefficient for UAV networks. Even the communication protocols designed for vehicular networks do not accommodate the key issues of UAV networks like their limited communication range, limited energy, limited processing power, faster speed, and structure-free mobility. \myblue{In \cite{dronestonextlevel}, distributed UAV networks are thoroughly studied with reviewing FANET structures and utilized networking protocols to highlight the current networking challenges and issues.}
\myblue{They conclude that} optimal routing and maintaining connectivity remain as two key challenging issues \cite{gupta2015survey}, \ar{which have been studied in several subsequent papers}.

\myblue{Machine Learning (ML) algorithms support efficient parameter estimation and interactive decision-making in wireless networks by learning from data and past experience. A review of using ML methods for different aspects of wireless networking is provided in \cite{ML30}. Also, ML algorithms facilitate the abstraction of different networking tasks from network topology prediction and channel status estimation by enabling embedded learnability features. 
For instance, Reinforcement Learning (RL)-based routing involves finding the most convenient path for any source-destination pair through the network based on different optimization criteria without directly monitoring and incorporating the network topology.} 
\myblue{Typically, the nodes' local information is used to take optimal communication decisions to minimize the overall energy consumption and enhance network connectivity \cite{rlintro}.} Reinforcement learning has also been used for other aspects of UAV networks, including spectrum management \cite{rlali1} and intelligent jamming defense \cite{rljamming}.

\subsection{Prior work}

RL-based routing was first introduced in \cite{q_routing}, where Q-routing is utilized as an application of packet routing based on Q-learning. This method demonstrated superior performance, compared to a non-adaptive algorithm based on pre-computed shortest paths \cite{network_qrouting}.
The essence of Q-routing is gauging the impact of routing strategies on a desired performance metric by investigating different paths in the \textit{exploration} phase and using the discovered best paths in the \textit{exploitation} phase. 
Exploration imposes an overhead to the system but is critical for finding newly emerged optimal paths, especially when the network topology undergoes substantial changes. An essential challenge is to solve the trade-off between the exploration and exploitation rate constantly to accommodate the level of dynamicity of the network topology.
An extension of the conventional Q-routing, known as Predictive Q-routing (PQ-routing) \cite{pred_q_routing}, attempted to address this issue and fine-tunes the routing policies under low network loads. Their approach was based on learning and storing new optimal policies under decreasing load conditions and reusing the best learned experiences by predicting the traffic trend. 
Their idea was to re-investigate the paths that remain unused for a while due to the congestion-related delays. They \ar{considered probing frequency as} an adjustable parameter that should be tuned based on the path recovery rate estimate. Results showed that PQ-routing outperformed the Q-routing in terms of both learning speed and adaptability. However, PQ-routing requires large memory for the recovery rate estimation. 
Also, it \ar{was not accurate in estimating} the recovery rate under varying topology change rates (e.g., when nodes start moving faster \ar{or slower}). Furthermore, this method only works \ar{when delays arise} from the queuing congestion, and not \ar{from} the network topology change. 

Another modification of the conventional Q-routing is Dual Reinforcement Q-routing (DRQ-Routing) \cite{drq_routing}. 
Their idea was to use forward and backward explorations by the sender and receiver of each communication hop by appending information to the \ar{data} packets they receive from their neighbors. Simulation results prove that this method learns the optimal policy more than twice faster than the standard Q-routing.
A comparative analysis of learning-based routing algorithms is provided in \cite{comparison}, where the performance of the self-adaptive Q-routing and dual reinforcement Q-routing algorithms \ar{is} compared against the conventional shortest path algorithms. \ar{Their results} showed that the Q-Learning approach \ar{outperforms} the traditional non-adaptive approaches \ar{when increasing traffic causes more frequent node and link failures}.
However, Q-routing does not always guarantee \ar{finding} 
the shortest path and does not explore multiple forwarding options in parallel. 

Two improved versions of Q-routing, namely Credence-based Q-routing (CrQ-Routing) and Probabilistic Credence-based Q-routing (PCrQ-Routing), are proposed in \cite{crq_routing} to capture the traffic congestion dynamically and to improve the learning process to select less congested paths. CrQ-Routing uses variable learning rates based on the inferred confidence to make the Q-value updates more efficient. Q-value updates are monitored to assess the freshness and accuracy of the measurements. A higher learning rate is used for the old updates, and \ar{a} lower learning rate \ar{is used} for the newer updates. Probabilistic Credence-based Q-routing (PCrQ-Routing) \ar{takes a random selection approach} to select a less congested path. 
\ar{This algorithm reverts} 
back to the optimal selection policy when the utilized confidence approach does not learn \ar{the} traffic load accurately and takes decision merely based on the information freshness. Both methods adapt to the current network conditions much faster than the conventional Q-routing.

Another technique to accelerate the learning speed of conventional Q-routing is the \textit{full-echo} approach \ar{introduced in} \cite{q_routing}. In conventional Q-routing, each node only updates the Q-values for \ar{the selected next-node (i.e., the best neighbor)}, whereas in the \textit{full-echo} routing, a node gets each neighbor's estimate of the total time to the destination \ar{to update} the Q-values accordingly for each of the neighbors. A more recent work added adaptive learning rates to the \textit{full-echo} Q-routing to improve the exploration performance \cite{adap_q_routing}. The adaptive \textit{full-echo} Q-routing uses two types of learning rates, one fixed (basic) rate for the neighbor to whom the packet is sent and another one (additional) for the rest of the neighbors. 
The additional learning rate changes dynamically according to the estimated average delivery time and allows to explore other possible routes. 
\ar{Their results show} that this technique reduces the oscillations of the \textit{full-echo} Q-routing for high-load scenarios. An extension of this work, Adaptive Q-routing with Random Echo and Route Memory (AQRERM) is introduced in \cite{aqrerm}, which improves the performance of the baseline method in terms of the overshoot and settling time of the learning process, as well as the learning stability.

\myblue{Recently, more advanced routing algorithms are proposed to extend the baseline Q-Routing into more complex scenarios with enhanced performance. Three successful algorithms include Poisson's probability-based Q-Routing (PBQ-Routing) \cite{pbq-routing}, Delayed Q-Routing (DQ-routing) \cite{dq-routing} and Qos-aware Q-Routing (Q$^2$-Routing) \cite{q2routing}.} PBQ-Routing uses forwarding probability and Poisson's probability for decision making and controlling transmission energy for intermittently connected networks. \ar{The results of this work} show that the delivery probability of this method is almost twice bigger than that of \ar{the} standard Q-routing while reducing the overhead ratio to half. DQ-Routing updates Q-values with random delays to reduce their overestimation and improve learning rate. Q$^2$-Routing includes a variable learning rate based on \ar{the amount of variation in Q-values while meeting the Quality of Service (QoS) requirements for the offered traffic.}

Unfortunately, \ar{all of} these Q-routing methods suffer from \ar{one or more} weaknesses, \ar{which negatively affect} their performance in extremely dynamic UAV networks. First, the methods that require large memories to store the history of the Q-values \ar{or the history of experienced delay (or other performance metrics) for each decision} become prohibitively restrictive in memory-constrained drones. \ar{The second issue is the Q-routing protocols' incapability in adapting their learning rate to varying network dynamicity rates.}  

\myblue{Another class of routing protocols is position-based algorithms, which use different tracking systems to exploit and monitor other nodes' positions. These algorithms select their path based on their inferred location information, i.e., by 
directly incorporating nodes' mobility into \ar{the path selection mechanism} to \ar{address} connectivity issues, which includes greedy distance-based methods \cite{khaledi2018greedy}. 
A complete review of position-based routing protocols when applied to 3D networks is presented in \cite{geographicrouting}. However, these methods require sophisticated tracking systems and large computation overhead for timely estimation of the location of all surrounding nodes \cite{razi2019optimal}. Indeed, these methods do not use the power of ML methods to indirectly learn the influence of the position information on performance metrics and realize more intelligent and independent decision making for routing protocols.} 
\myblue{The above-mentioned facts led the researchers to use RL-based routing methods. Some of the RL-based routing algorithms include predictive ad-hoc routing fueled by reinforcement learning and trajectory knowledge (PARRoT) \cite{parrot}, adaptive and reliable routing protocol with deep reinforcement learning (ARdeep) \cite{ardeep}, traffic-aware Q-network enhanced routing protocol based on GPSR (TQNGPSR) \cite{TQNGPSR} and Q-learning based multi-objective optimization routing protocol (QMR) \cite{QMR}, as well as RL-based routing protocols that use fuzzy logic for decision-making \cite{fuzzy1, fuzzy2, fuzzy3}. Although the proposed provide elegant solutions for routing, they do not seem to be practical with current drone technology due to their high computational complexity. Also, they perform poorly in adapting to abrupt network changes.}

\subsection{Contributions}

Noting the shortcomings of the existing RL-based routing protocols, in this paper, we introduce a promising Q-learning-based routing protocol that is suitable for highly dynamic UAV networks. Low complexity, low overhead \ar{requirements}, local forwarding decision, and no \ar{need for} initial route setup are some of the key characteristics of the proposed method to improve the probability of \ar{successful} packet delivery in UAV networks. 
The major contributions of this paper are summarized as follows:

\begin{itemize}

    \item First, we introduce \ar{a trajectory creation approach, which uses} a piece-wise linear mobility model to produce node trajectories. It consists of a hierarchical generative model that defines random parameters for each UAV class, representing each node's motion profile. This model is suitable for \ar{UAV networks with heterogeneous mobility levels (e.g., networks of quadcopters, mini-drones, and fixed-wing UAVs)} since it is easy to infer the motion profile of each node by sampling its motion trajectory.

    \item Secondly, we propose a full-echo Q-routing with an adaptive learning rate controlled by Simulated Annealing (SA) optimization, where the \textit{temperature} parameter captures the influence of the nodes' mobility on the update rates of Q-value. The soft variation of the exploration rate \ar{with the re-initiation feature} not only optimizes the exploration rate, but also accommodates abrupt changes in the network dynamicity. The criteria we used \ar{for path selection minimizes} the packet transmission energy.

    \item Lastly, we performed extensive simulations to assess the performance \ar{and the complexity} of the proposed algorithms, compared with previous Q-routing algorithms \ar{and Q-routing with other heuristic optimization methods} using different network scenarios. 
    The quantitative results confirm a considerable reduction in energy consumption and an increase in the packet delivery rate for the proposed algorithm.
    
\end{itemize}

\myblue{The rest of this paper is organized as follows. In section \ref{sec:systemmodel}, we introduce the system and the utilized mobility model. In section \ref{sec:routingprotocol}, we describe the proposed fully-echoed Q-routing algorithm equipped with Simulated Annealing inference. 
This section also includes a computational complexity analysis of the proposed method in comparison with similar methods. 
In section \ref{sec:results}, we present the simulation results with quantitative analysis. Finally, the main findings of this work are reviewed in section \ref{sec:conclusion}.}


\section{System model}   \label{sec:systemmodel}
\begin{figure}[h]
    \centering
	\includegraphics[width=0.85\columnwidth]{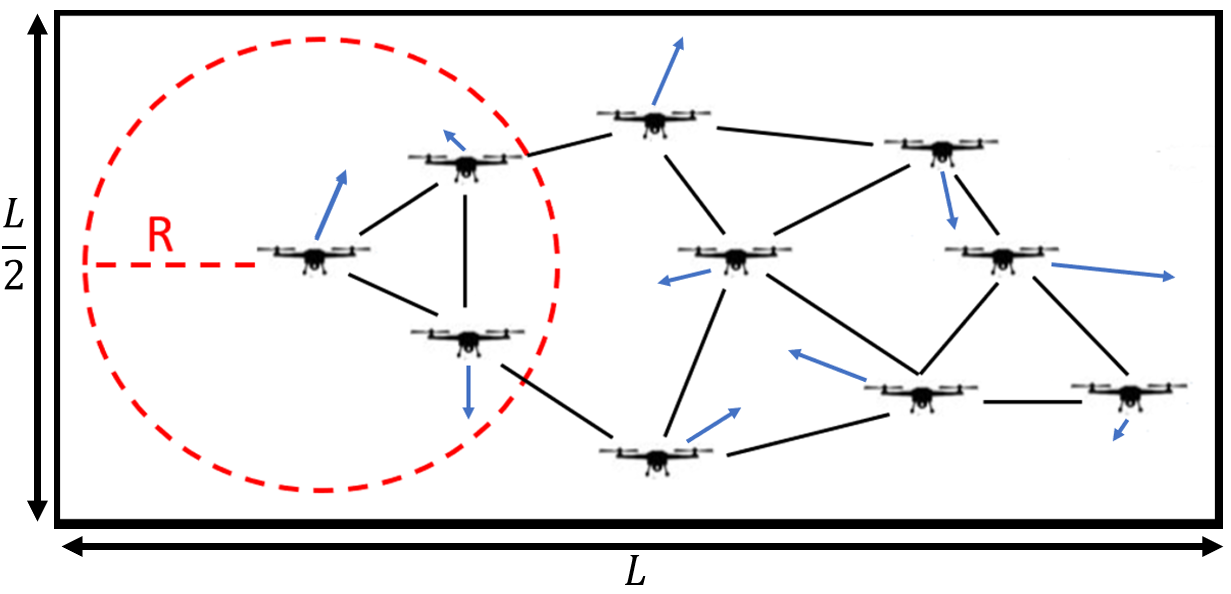}
    \caption{Network topology \ar{with dynamic contact graph based on the nodes' communication range.}}
     \label{fig:system_model}
\end{figure}

We consider a wireless mesh network composed of $N$ nodes $\mathcal{N}=\{n_1,n_2,\dots,n_N\}$ distributed uniformly in a rectangular area of an arbitrary size $L\times L/2$, as depicted in Figure \ref{fig:system_model}. The communication range of each UAV is represented by a circular area of radius $R$. Therefore, the set of the neighbors for node $n_i$ is defined as:
\begin{align} \label{eq:network}
S_i(t)= \Big\{n_j\in \mathcal{N}: d_{ij}(t)\leq R\Big\} 
\end{align}

\noindent where $d_{ij}(t)=\sqrt[]{(x_i(t)-x_j(t))^2+(y_i(t)-y_j(t))^2}$ is the Euclidean distance between nodes $n_i$ and $n_j$ at time $t$. This realizes a dynamic contact graph where there exists a communication link between any pair of nodes within a given distance. 

\begin{figure}[h]
    \centering
	\includegraphics[width=0.6\columnwidth]{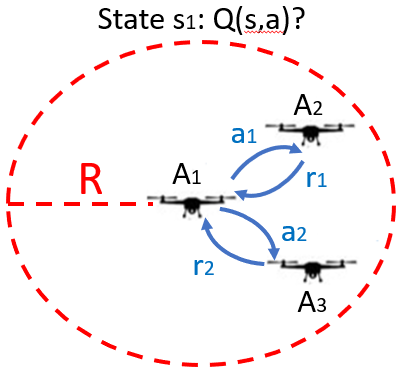}
    \caption{Illustration of the RL-based routing.}
     \label{fig:rl_routing}
\end{figure}

Here, we use Q-learning, a variant of model-free reinforcement learning that enables optimal decision making by evaluating the rewards of actions in an uncertain or unknown environment with no central supervisor \cite{reinforcementlearning}. Q-learning is a \ar{variant of the} reinforcement learning algorithm, which provides agents $A_i$ with the capability of directly learning the consequences of their actions $a_i$ (which node to send the packet to) when they are at specific states $s_i$ (e.g., location, traffic load, etc.) in terms of \ar{the} achieved reward $r_i$. The reward is defined as the reduction \ar{of a} desired performance metric, i.e., the transmission energy from the source to the destination, achieved by the action. The concept of reinforcement learning for optimized routing is shown in Figure \ref{fig:rl_routing}. In the initial environment represented by state $s_1$, node/agent $A_1$ has two candidate \ar{neighbors} $A_2$ and $A_3$ to send its packet. The spirit of RL-based routing is selecting one of the actions $a_1$ or $a_2$ \ar{based on the reward expected for each} action $a$ at state $s$, defined as $Q(s,a)$. Once we establish an optimal forwarding decision policy ($a_1$ or $a_2$), the agent $A_1$ obtains an immediate reward from the environment, $r_1$ or $r_2$, respectively. Then, it transits to state $s_2$, where new decisions are made based on the new environmental conditions and the learned policy in terms of actions-rewards relations. The end goal is to find an optimal policy in which the cumulative reward over time is maximized by assigning optimal actions to each state.

The idea behind the full-echo \ar{routing} equipped with SA optimization is to go one step beyond the conventional Q-routing \ar{paradigm} and adapt the learning rate of the algorithm based on the rate of topology change, implied by the node velocities. The goal is to keep the overall energy consumption of the network at the minimum level possible while minimizing the packet drop rate for selecting unstable links.

\ar{Our approach is to regulate the exploration rate of Q-routing to identify potentially new optimal paths while not overusing exploration time}. 
Some prior works intend to directly incorporate the predicted network topology into the routing algorithm \cite{rovira2017predictive,razi2018predictive}. Despite their near-optimal performance, these methods require accurate tracking systems that can be restrictive in real scenarios. 
Here, we use simulating annealing optimization to adjust the exploration rate by indirectly learning the impact of node mobility on the communication energy consumption. 

\begin{figure}[t]
	\centering
	\includegraphics[width=.8\columnwidth]{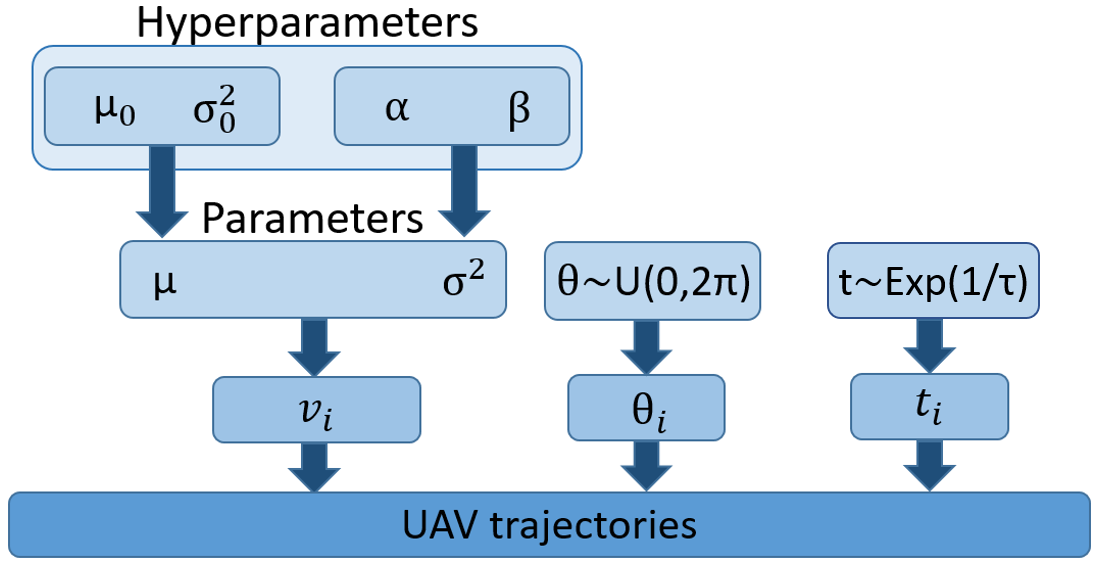}
	\caption{\ar{A hierarchical generative model} used to generate \ar{class-specific motion} trajectories.}
    \label{fig:trajectoriesgen}
\end{figure}

\subsection{Piece-wise linear mobility}
To \ar{emulate } networks with heterogeneous mobility parameters, we use a parametric generative model to produce node trajectories. 
The mobility model consists of piece-wise linear motions over time intervals whose duration is exponentially distributed as $t_i \sim Exp(1/\tau)$\footnote{This model can be viewed as a waypoint model with linear motions between the waypoints that has the flexibility of generating arbitrary trajectories when the interval between \ar{selected consecutive waypoints} is small enough.}.
During each segment $t_i$, we use a constant velocity $v_i$ and direction $\theta_i$ that vary \ar{for the next segment.} 

A hierarchical generative model is used to produce the parameters that ultimately define the motion trajectories (Figure \ref{fig:trajectoriesgen}). Consider each node moves with a random but fixed velocity $v_i$ in a random direction $\theta_i$, at each interval $t_i$. The velocity $v_i$ is a Random Variable (RV) with Gaussian distribution $v_i \sim \mathcal{N}(\mu_{vi},\,\sigma^{2}_{vi})$. The direction $\theta_i$ follows a uniform distribution $\theta_i \sim \mathcal{U}(0, 2\pi)$. \ar{These distributions are used to simulate the worst-case scenario following prior works in \cite{priordistribution}. More specifically, Gaussian distribution maximizes the entropy under limited energy, hence is appropriate for creating the most unpredictable node velocities. Likewise, the uniform distribution is the most uninformative distribution that maximizes entropy for RVs with limited range, like the direction $\theta_i$, which is limited to the $[0,2\pi]$ range.
However, for the sake of completeness and to ensure that the results are generic, we examined our routing protocol using other widely adopted mobility models for UAV networks, including random waypoint \cite{rwp}, Gauss-Markov Mobility Model \cite{gmmm}, and Paparazzi mobility model \cite{paparazzi}.}

\ar{Another advantage of using a segment-wise mobility model with a} symmetric distribution for $\theta$, is to prevent the network from falling apart, as occurs for networks with linear motions at random directions. 

The triplet ($t_i,\theta_i,v_i$) forms the model parameters. To accommodate nodes with different velocity profiles, the mean $\mu_{vi}$ and variance $\sigma^{2}_{vi}$ of $v_i$ are considered RVs controlled by hyper-parameters. In particular, we have $\mu_{vi} \sim \mathcal{N}(\mu_0, \sigma^{2}_0)$ and $\sigma^{2}_{vi} \sim Inv-Gamma(\alpha, \beta)$, an inverse Gamma distribution with shape $\alpha$ and rate $\beta$. The hyper-parameters $\alpha$, $\beta$, $\mu_0$ and $\sigma^{2}_0$ are the same among all nodes, and \ar{are used to obtain} node-specific model parameters $\mu_i$ and $\sigma^{2}_i$ for nodes $n_i=(1,2,\dots,N)$. The model parameters represent the motion profile of each node based on its class, and remain constant throughout the operation time. This hierarchical modeling with conjugate priors for the model parameters facilitates deriving closed-form posterior and predictive probabilities for the model parameters, to easily infer the motion profile of each node by sampling its motion trajectory. We use this model to generate the motion trajectories for all network nodes \ar{and to create} the dynamic time-varying contact graph of the network using (\ref{eq:network}).


\myblue{Another assumption that we considered is} that the motion intervals are long enough to let the learning algorithm converge to an optimal solution with a reasonable learning rate. This assumption is relevant since the employed learning algorithm requires only around fifty transmission rounds for a full convergence that remains in the millisecond range while the motion changes for UAVs occur in the second range, if not minutes. However, the algorithm is flexible enough to recognize the change points and re-adapt to the new velocities with no human intervention, \ar{noting the re-initiation process at the beginning of each segment}. The learning rate is determined by the temperature parameter $T$ for the utilized SA algorithm, which quickly adapts to the network nodes' average velocities during each interval. This adaptation is realized without the need for directly inferring the node velocities using sophisticated tracking systems.

\section{Routing protocol} \label{sec:routingprotocol}


The proposed routing protocol enables the nodes to make packet forwarding decisions based on their local experience with the ultimate goal of minimizing the end-to-end transmission energy. No prior information is required about the network nodes' mobility and traffic load distribution across the network.  

We first review conventional Q-routing \cite{q_routing} and then indicate the modifications we made to develop our proposed method, namely the \textit{fully-echoed Q-routing} with an adaptive learning rate using the inferred SA parameters. The Q-value $Q_x(d,y)$ is defined as the time-span it takes for node $x$ to deliver a packet to the destination node $d$ through neighbor node $y$. Then, after sending the packet from node $x$ to node $y$, node $y$ estimates the remaining time for the trip $t_{y \rightarrow  d}$, defined by:

\begin{equation} \label{t}
t_{y\rightarrow d} = \min_{z \in \mathcal{S}_y(t)} Q_y (d,z),
\end{equation}
where $\mathcal{S}_y(t)$ is the set of the neighbors of $y$ at current time $t$.
Next, from the information that node $x$ receives, we can update $Q_x(d,y)$ to:
\begin{equation} \label{qx}
Q_x(d,y)_\text{new} = Q_x(d,y)_\text{old} + \eta \big( q+s+t - Q_x(d,y)_\text{old}\big),
\end{equation}
where $q$ is the waiting time for node $x$ and $s$ is the transmission time from node $x$ to node $y$. Also, $\eta$ is an adjustable learning rate. 
To improve the learning speed, using the full echo Q-routing with adaptive learning rates, we update Q-tables for all neighbors by sending estimation packets to the neighbors. We define two learning rates: basic ($\eta$) and additional ($\eta_2$). Each node updates its Q-table using $\eta$ if it refers to the neighboring node \ar{to which} we sent the packet, and using $\eta_2$ otherwise. The basic learning rate ($\eta$) is fixed; however, the additional learning rate ($\eta_2$) is updated at each step using
\begin{equation} \label{eta_2}
\eta_2 = \frac{T_{est}}{T_{max}}\cdot \eta \cdot k
\end{equation}
where $T_{est}$ is the estimate of the average delivery time and $T_{max}$ is the estimate of the maximum average delivery time. Also, $k$ is a predefined parameter to be tuned by the experiments for optimal performance.

The exploration rate is controlled by the SA algorithm \cite{sima_qlearning}, as a natural optimization choice. 
\myblue{The reason for selecting SA as the optimization algorithm compared to other heuristic optimization methods such as Gradient Descent (GD), Genetic Algorithm (GA), and Particle Swarm Optimization (PSO) is that SA's naturally embedded property of starting from more aggressive exploration rates (at high temperatures) and leaning gradually toward more conservative decisions over time by cooling down the temperature parameter T, makes it desirable for segment-wise routing decisions. This feature accommodates dynamic topology with abrupt changes.} 
\myblue{The temperature $T$ changes exponentially from $T=(k_{max}/k)$ to $T=1$}, where $k$ is the iteration and $k_{max}$ is the \ar{maximum allowable number} of iterations for exploration. 
Here, we control the temperature cooling \ar{based on} the velocities of the network nodes captured by the changes in the selected links' performance. Once a velocity change is detected (at the beginning of an interval), the temperature automatically is increased to the highest value and cools down gradually during the interval. 

The summary of the operation of \ar{one full cycle} of the SA optimization is provided in Algorithm \ref{alg:simulatedannealing}, where we have:
\begin{align}\label{eq:1}
\begin{cases}
P(a_p,a_r,T)=1, & \text{if} \,a_r<a_p,\\
P(a_p,a_r,T)=e^{\frac{-(a_r-a_p)}{T}}, &\text{otherwise}.
\end{cases}
\end{align}

Here, $P(a_p,a_r,T)$ acts as the \ar{exploration probability}. $P=1$ means that the random action $a_r$ is better than the previously identified best action $a_p$, and we select $a_r$. Otherwise, we select the next node based on this probability by taking a random action $a_r$ with probability $P(a_p,a_r,T)$ and following the best action with probability $1-P(a_p,a_r,T)$.
Here, \ar{a} random action $a_r$ means the next node is selected uniformly among the available neighbor nodes. 

\begin{algorithm}
 \caption{Q-routing table update using SA}
 \ar{
 \begin{algorithmic}[1]
  \STATE \textit{Initialize Q-values ($Q_x(d,y)$) for all neighbors $y$;}
  \STATE \textit{Initialize $f, \eta$;}
  \FOR {$k=1$ to $k_{max}$}
  \STATE T $\leftarrow k_{max}/k$
  \STATE T $\leftarrow T \times f$
  \STATE Select action $a_r$ uniformly among neighbors;
  \STATE Select action $a_p$ according to learned Q-values;
  \STATE $a \leftarrow a_p$;
  \STATE generate random variable $r\sim \mathcal{U}[0,1]$
  \IF {($P(a_p,a_r,T) \geq r$)}
  \STATE $a\gets a_r$
  \ENDIF
  \STATE \textbf{Execute} action $a$
  \STATE \textbf{Evaluate} $f$ using (\ref{eq:factor})
  \STATE \textbf{Update} Q-value: \\
  $\Rightarrow$ for selected neighbor:\\ 
    $Q_x(d,y)_\text{new} = Q_x(d,y)_\text{old} + \eta \big( q+s+t - Q_x(d,y)_\text{old}\big)$ \\
   $\Rightarrow$ for the rest of the neighbors:\\
    $Q_x(d,y)_\text{new} = Q_x(d,y)_\text{old} + \eta_2 \big( q+s+t - Q_x(d,y)_\text{old}\big)$ \\
      (where $\eta_2 = \frac{T_{est}}{T_{max}}\cdot \eta \cdot k$) \\
  \ENDFOR
 \end{algorithmic}
 }
\label{alg:simulatedannealing}
 \end{algorithm}



\ar{The dynamicity of the network is indirectly inferred by the change of the consumed energy (or equivalently the variation of Q-values) during the last H iterations. More specifically, we define parameter $f$ as:
\begin{align}
\label{eq:factor}
f=\gamma |\Delta E|=\frac{1}{H}\sum_{i=1}^{H}|E_{k+1-i}-E_{k-i}|,
\end{align}
where $H$ is the length of history, to be selected based on the velocity of the nodes and the length of each interval to identify significant variations. Here, we choose $H=10$. Parameter $\gamma$ is a scaling parameter to map the energy variation into the $[0.5,10]$ range. The parameter $f$ is used to regulate the temperature cooling in the SA algorithm by scaling the temperature parameter $T$ depending on how fast the chosen path's energy changes over time.}




\subsection{Computation complexity} \label{sec:computational}

\myblue{Here, we include a comparative analysis of the computation complexity of the proposed method, compared to other state-of-art routing algorithms, including ad hoc on-demand distance vector (AODV) \cite{AODV}, optimized link state routing (OLSR) \cite{OLSR}, destination sequenced distance vector (DSDV) \cite{DSDV}, dynamic source routing (DSR) \cite{DSR}, greedy perimeter stateless routing  (GPSR) \cite{GPSR}, as well as the conventional Q-routing \cite{q_routing}. 
Computation complexity is defined as the number of operations required to execute one round of an algorithm. 
Results are shown in table \ref{complexity}, where $N$ represents the number of nodes or network size. We can see that the proactive routing protocols (e.g., OLSR, DSDV) and reactive routing protocols (e.g., AODV, DSR) have low and average complexity, respectively. Position-based routing protocols (e.g., GPSR) and learning-based routing protocol (e.g., Q-routing and the proposed method) have a higher complexity. \ar{Our complexity is quadratic in $N$, which is higher than proactive and reactive methods, but still affordable for reasonable network sizes.} It has been shown that learning-based decentralized methods that adapt to dynamic networks without the need for global knowledge and long route setup are more suitable for UAV networks since they eliminate the need for costly and sophisticated positioning methods. Our proposed method constantly adapts to both minor or abrupt changes, leading to a higher packet delivery ratio and energy efficiency, as well as retaining maximal connectivity. For this reason, the higher complexity, compared to proactive and reactive routing protocols, is justified by the increase in routing performance. Lastly, we can state that our method has lower complexity with respect to popular topology-aware routing protocols while providing better results.}

\ar{
\begin{table}[h]
\centering 
\caption{\small Complexity for different routing protocols}
\label{complexity}
{\footnotesize
\begin{tabular}{|c|c|}
\hline
\rowcolor{gray}
\textbf{Routing protocol} & \textbf{Computation complexity} \\
\hline
AODV \cite{AODV} & $O(2N)$ \\
\hline
OLSR \cite{OLSR} & $O(N)$ \\
\hline
DSDV \cite{DSDV} & $O(N)$  \\
\hline
DSR \cite{DSR} & $O(2N)$ \\
\hline
GPSR \cite{GPSR} & $O(N^3)$  \\
\hline
Q-Routing \cite{q_routing} & $O(N^2)$  \\
\hline
\rowcolor{lightgray}
Proposed & \textbf{$O(N^2)$} \\
\hline
\end{tabular}
}
\end{table}
}

\subsection{Memory requirements}

\myblue{In this section, we study the memory requirements and the storage efficiency for the proposed routing protocol. 
The proposed routing protocol needs Q-table resources for each node, with a short history to consider the adaptability to the network state at each step. In section \ref{sec:routingprotocol}, we defined the exploration-exploitation approach that considers the history of the last 10 packets to reevaluate the exploration rate. Consequently, if we study computational complexity or memory requirements for Q-routing-based routing protocols, they suffer from the curse of dimensionality. The Q-table grows at $O(N^2)$ with the number of nodes.}

\myblue{However, compared to standard RL or deep-learning-based algorithms such as \cite{parrot, ardeep, QGeo, QNGPSR}, the memory requirements for our method are relatively low. RL algorithms require large memory for relatively large state spaces and \ar{close-to-one reward discount rate}. Deep learning algorithms are also computationally extensive and require large memory to store training samples and the history of reward-action pairs. Therefore, their learning phase may take much longer than for the proposed methods. Consequently, it is fair to state that the computational complexity of our Fully-echoed Q-routing with SA inference method is doable with on-board memory capabilities in UAVs, in contrast to other more complex and memory-needy solutions that provide similar outcomes.}

\subsection{Overhead analysis}

\myblue{Overhead is defined as the number of additional routing packets sent for route discovery, establishment, and maintenance. An advantage of our method is that we do not use exploration packets (like sending periodic \textit{hello} packets in proactive routing protocols such as OLSR algorithm \cite{OLSR}) to find optimal paths; rather, it is learned by monitoring \textit{data} packets in the exploration phase. Our method addresses the essential trade-off between exploration and exploitation based on the network's behavior. During the exploration phase, we learn the network's state by studying all neighbors' behavior, and these suboptimal transmissions can be considered exploration overhead. In the exploitation phase, the overhead is considerably low since only the best identified paths are utilized. Since our method addresses this trade-off using the SA optimization module based on the network's dynamicity level, the incurred overhead is much lower than other routing protocols.}

\myblue{We analyze how overhead impacts our routing protocol by investigating if the incurred overhead correlates with a better learning state or not, and what is the rate of taking non-optimal decisions. We expect to find a trade-off between the exploration rate (that brings more overhead) and the knowledge of the network. In table \ref{overhead}, we study the effect of exploration rate with reaching the optimal solution by counting the number of packets we send to reach the optimal solution. We can observe that our proposed adaptive method gives the best result in terms of optimality of the path selected and the number of packets needed to reach that solution. If we fix the exploration rate to realize a low or medium exploration rate, the algorithm converges quickly to the final path, with 2 and 14 packets, respectively. However, the algorithm may not converge to an optimal solution since it may select some of the intermediate nodes inaccurately. On the other hand, if a fixed high exploration rate is selected, the optimal solution can be found faster, but it takes an average of 41 packets to converge. The proposed routing protocol finds the best path faster than fixed exploration rates (with an average of 4 packets) and quickly converges to the best solution. This means that the proposed algorithm with an adaptive exploration policy outperforms the fixed-rate algorithm both in terms of fast convergence and the optimality of the solution.} 

\begin{table}[h]
\centering 
\caption{\small Effect of exploration rate in finding optimal solution.}
\label{overhead}
{\footnotesize
\begin{tabular}{|c|c|c|} 
\hline
\textbf{Exploration rate} & \textbf{Optimal selection} & \textbf{Packets until converged} \\
\hline
Low & No & 2 \\
\hline
Medium & No & 14 \\
\hline
High & Yes & 41 \\
\hline
\specialcell{\textbf{Proposed}\\\textbf{(adaptive)}} & \textbf{Yes} & \textbf{4} \\
\hline
\end{tabular}
}
\end{table}

\myblue{Concluding, we observe how the adaptive protocol has a lower overhead than high exploration methods, as it needs fewer packets to learn the state of the network.}

\section{Simulation results}   \label{sec:results}

To assess the performance of the proposed method, we compare it against the state-of-the-art Q-routing algorithms discussed in section \ref{sec:introduction}, including (i) Random Exploration-Exploitation Routing (REE-Routing), (ii) Probabilistic Exploration Routing (PE-Routing), (iii) Conventional Q-routing \cite{q_routing}, (iv) Adaptive learning rates Full-Echo Q-Routing (AFEQ-Routing) \cite{adap_q_routing}, and (v) Simulated Annealing based Q-routing (SAHQ-Routing) \cite{simann_q_routing}. 
Methods (i) and (ii) are simulated for the sake of comparison only.  
\myblue{We compare only against other Q-routing-based routing protocols, as previous works have shown that the learning-based algorithms outperform AODV, OLSR, and GPSR, among other well-known routing protocols \cite{parrot, learningrouting}.}
We simulate different network scenarios using the piece-wise linear mobility (Figure \ref{fig:trajectoriesgen}). \myblue{This model uses the entropy-maximizing Gaussian distribution for the seed and the most uninformative uniform distribution for the direction to simulate the worst-case scenario, although other mobility models for UAV networks could be used.} To realize a fair comparison, we use the same set of trajectories to test different algorithms.

The first set of comparative results is presented in Table \ref{table_improvement}, in terms of the end-to-end transmission energy. We simulate networks with different sizes ($N=10$, $N=20$) and three velocity profiles of slow-speed ($\mu_0=10, \sigma^{2}_0=2.5$), medium-speed ($\mu_0=25, \sigma^{2}_0=5$), and fast-speed ($\mu_0=50, \sigma^{2}_0=10$) with $\alpha=5, \beta=1$ for all scenarios. The communication range is $R=7500$ meters. The proposed algorithm considerably improves upon the performance of all algorithms consistently by reducing the average energy consumption. The achieved gain ranges from $7\%$ to $82\%$ depending on the reference method and the utilized network parameters. We observe that our method offers higher gains for larger networks ($N=20$), and slower speeds ($\mu_0=10, \sigma^{2}_0=2.5$).



\begin{table}[h]
\centering
\caption{\small Comparative analysis: energy consumption \ar{of different routing algorithms including the proposed method under different} network sizes and velocity profiles.}
\label{table_improvement}
{\footnotesize
\hskip-0.5cm 
\begin{tabular}{|p{1.1cm} | p{0.8cm} | p{0.8cm} | p{0.8cm} | p{0.8cm}  |p{0.8cm} | p{0.8cm} |} 
\hline
& \multicolumn{3}{c|}{ $N$ = 10} & \multicolumn{3}{c|}{ $N$ = 20}\\
\hline
 & Slow & Medium & Fast & Slow & Medium & Fast\\
\hline
REER & 95.3 & 103.7 & 127.1 & 135.1 & 170.3 & 163.9  \\
\hline
PER & 146.3 & 146.5 & 149.5 & 269.0 & 289.0 & 246.6 \\
\hline
QR & 83.7 & 93.0 & 97.1 & 86.0 & 104.7 & 104.5 \\
\hline
AFEQR & 75.8 & 79.3 & 96.4 & 76.9 & 90.4 & 91.3 \\
\hline
SAHQR & 70.6 & 85.4 & 98.0 & 70.6 & 98.2 & 99.2 \\
\hline
Proposed & 65.6 & 70.6 & 87.0 & 47.8 & 76.0 & 82.5 \\
\hline
\textbf{Gain} & \textbf{7\%-55\%} & \textbf{11\%-52\%} & \textbf{10\%-42\%} & \textbf{32\%-82\%} & \textbf{16\%-74\%} & \textbf{10\%-67\%} \\ 
\hline
\end{tabular}
}
\end{table}

The communication range $R$ plays an essential role in the performance of routing algorithms. It not only influences the sparsity of the graph by affecting the node degrees but also affects the connectivity of the network under the utilized routing algorithm. 
More specifically, a packet drop occurs when a node has no neighbors within the communication range or the selected nodes go beyond the communication range while transmitting. 
The impact of the communication range on the packet drop rate is presented in Table \ref{table_packetdeliveryratio} for $R=5km$, $R=7.5km$, and $R=10km$.  \ar{It can be seen that} our method has a higher successful packet delivery rate compared to the baseline conventional Q-routing (QR), and performs better or almost equal to the rest of the methods. The gain in the packet delivery ratio can go as high as $264\%$ depending on the network size ($N$) and the communication range ($R$). For the lower communication \ar{range} ($R=5000m$), and smaller networks ($N=10$), the achieved gain (the reduction in packet drop rate) is higher. 

\begin{table}[h]
\centering
\caption{\small Packet delivery rate versus communication range for different algorithms.}
\label{table_packetdeliveryratio}
{\footnotesize
\begin{tabular}{|c|c|c|c|c|c|c|} 
\hline
& \multicolumn{2}{c|}{ $R$ = 5000} & \multicolumn{2}{c|}{ $R$ = 7500} & \multicolumn{2}{c|}{ $R$ = 10000}\\
\hline
& N=10 & N=20 & N=10 & N=20 & N=10 & N=20\\
\hline
REER & 24.5\% & 50.9\% & 82.0\% & 83.3\% & 94.2\% & 96.6\%  \\
\hline
PER & 18.6\% & 41.1\% & 70.1\% & 73.2\% & 86.5\% & 91.8\% \\
\hline
QR & 8.9\% & 38.9\% & 77.2\% & 85.4\% & 99.9\% & 99.9\% \\
\hline
AFEQR & 32.4\% & 70.1\% & 90.3\% & 95.9\% & 99.9\% & 99.9\% \\
\hline
SAHQR & 28.9\% & 61.4\% & 90.3\% & 96\% & 99.1\% & 99.9\% \\
\hline
Proposed & 32.4\% & 70.6\% & 90.3\% & 96.3\% & 100\% & 99.9\% \\
\hline
\textbf{Gain (up to)} & \textbf{264\%} & \textbf{81\%} & \textbf{29\%} & \textbf{32\%} & \textbf{16\%} & \textbf{9\%} \\ 
\hline
\end{tabular}
}
\end{table}

\begin{figure}[h]
\begin{subfigure}{0.95\columnwidth}
\centering
\includegraphics[width=0.95\columnwidth]{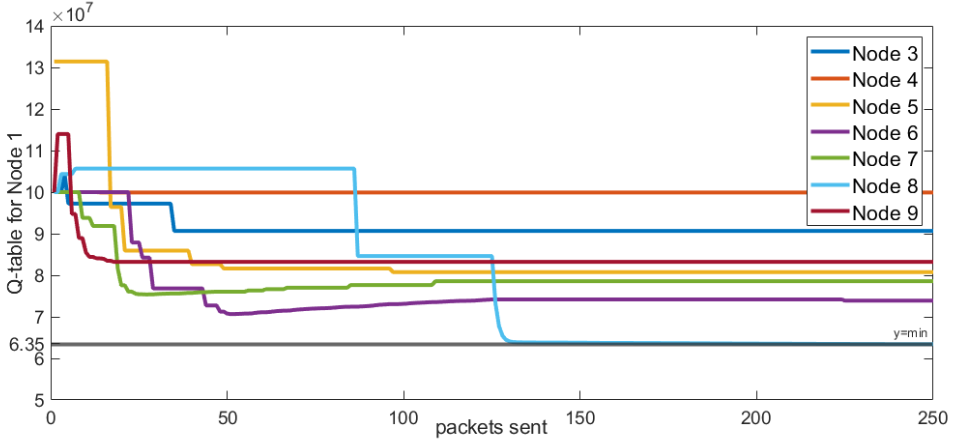}
\end{subfigure}
\begin{subfigure}{0.95\columnwidth}
\centering
\includegraphics[width=0.95\columnwidth]{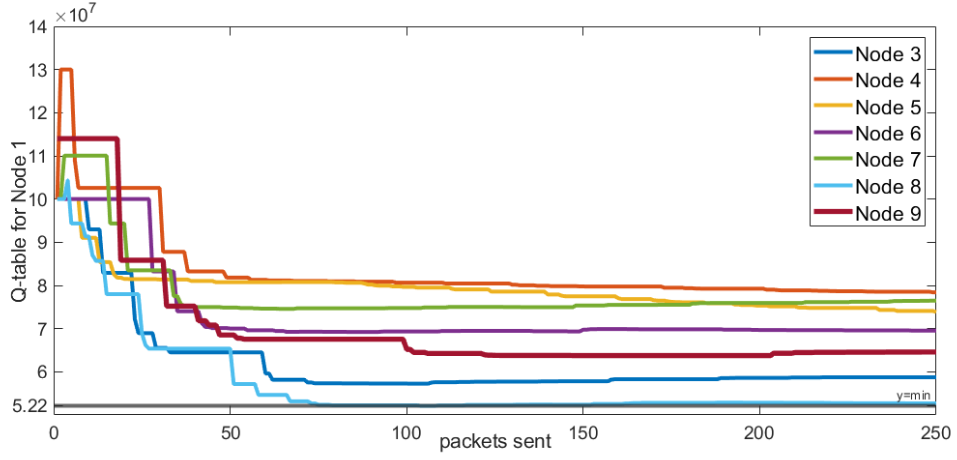}
\end{subfigure}
\begin{subfigure}{0.95\columnwidth}
\centering
\includegraphics[width=0.95\columnwidth]{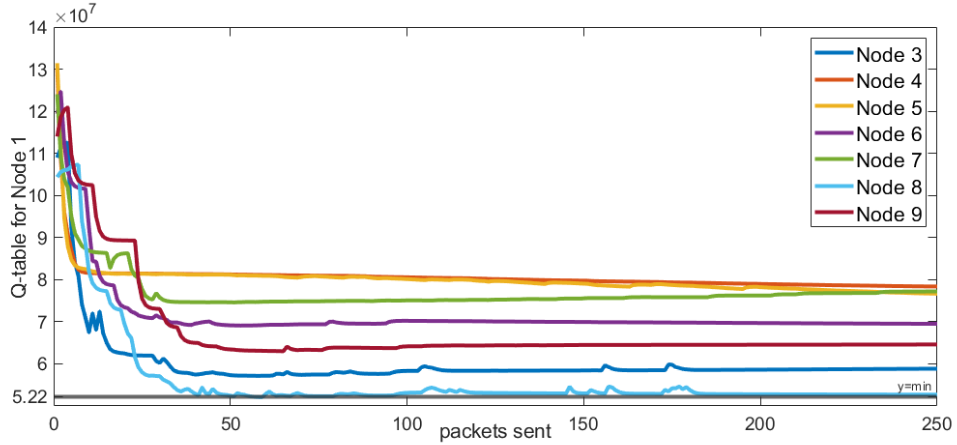}
\end{subfigure}
\caption{\small Q-table for an exemplary source node \ar{$n_1$ to send a packet to destination $n_2$ through any of its seven neighbors ($n_3,n_4,\dots,n_9$)} under different routing protocols including (top) SAHQ-Routing with non-adaptive parameter $T$ \cite{simann_q_routing}, (middle) high-exploration Q-Routing, and (bottom) the proposed method with flexible exploration rate.}
\label{fig:qtable}
\end{figure}

Next, we study the effect of the SA-based optimization on the evolution of the Q-values that represent the expected energy for a packet \ar{from the source node $n_1$ to the end destination $n_2$ through any of its seven neighbors ($n_3,n_4,\dots,n_9$).}
The Q-value evolution of the seven neighbors is depicted in Figure \ref{fig:qtable} for three algorithms\ar{, including} (i) SAHQ-Routing with no adaptability of T parameter, (ii) a Q-Routing with high exploration rate, and (iii) the proposed method. \myblue{Each line represents the evolution of the Q-values when each of the neighbors is selected as the next bode. It is shown how Q-values converge to the optimal solution over time when more packets are sent.} It can be seen that the SAHQ-Routing does identify the best next node (8) (after 125 rounds) but at a much lower rate compared to our method (after 40 rounds). Also, its recovered end-to-end path does not seem to be optimal (despite finding the best second node) since the minimum value in the Q-table is 63.5 Mega Joule (MJ) for SAHQ-Routing, compared to the 52.2 MJ for our method. A similar gain is achieved for our method, compared to the Q-Routing with high exploration. Nevertheless, our method converges to the optimal solution faster than the Q-Routing with high exploration (40 rounds in contrast to 70 rounds, respectively).
\myblue{In short, our solution converges to the optimal value at a faster rate than the other two competitor methods. This gain comes from the capability of our method in adapting to the dynamicity of the network.}


\begin{figure*}[h]
\begin{subfigure}{2\columnwidth}
\centering
\includegraphics[width=0.9\columnwidth]{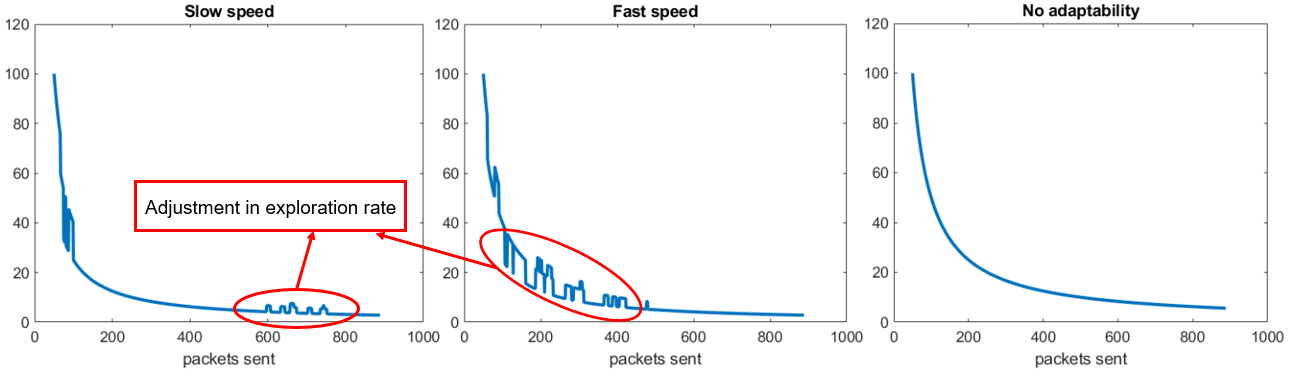}
\caption{}
\label{fig:temp_speed}
\end{subfigure}
\begin{subfigure}{2\columnwidth}
\centering
\includegraphics[width=0.9\columnwidth]{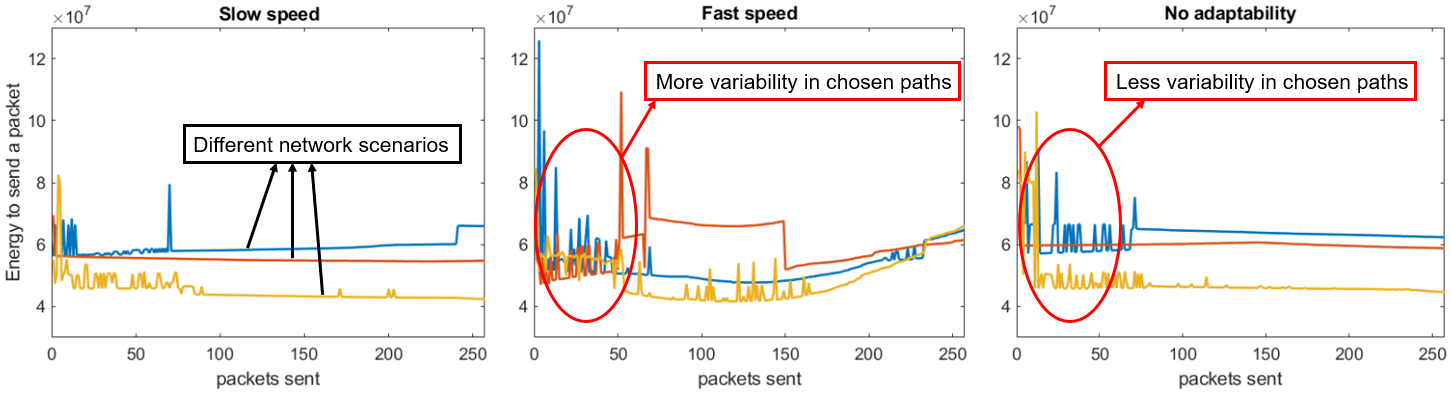}
\caption{}
\label{fig:ener_speed}
\end{subfigure}
\label{fig:temp_ener_speed}
\caption{a) The evolution of the temperature parameter $T$ for different network scenarios with (left) slow-speed, (center) fast-speed, and (right) protocol with no temperature adaptability \cite{simann_q_routing}; b) End-to-end energy consumption for different network scenarios.}
\end{figure*}

\begin{figure}[h]
	\centering
	\includegraphics[width=1\columnwidth]{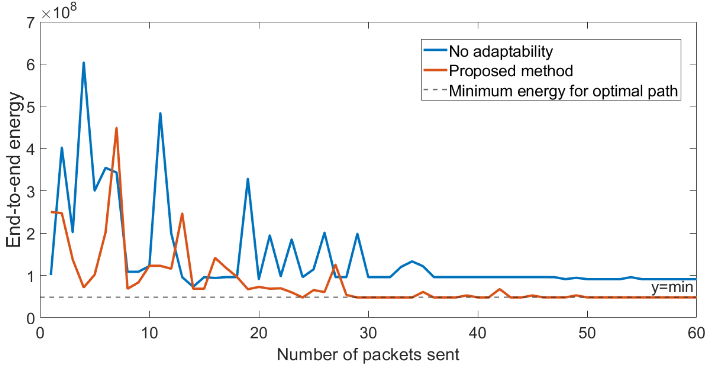}
	\caption{\small End-to-end energy of the proposed method vs non-adaptive temperature method to show the stability and reliability of the proposed method.}
    \label{fig:stability}
\end{figure}

Figure \ref{fig:temp_speed} (top row) illustrates the evolution of the temperature parameter $T$ over time. 
As described previously, $T$ is directly proportional to the exploration rate, and its natural behavior is exponential declining (i.e., $f(x)=e^{-x}$) iteration by iteration over time, as shown in the top-right most figure for the baseline method with \ar{non-adaptive $T$}. However, our method adjusts the temperature \ar{cooling} rate by examining the changes in the experienced consecutive Q-values. We observe that for a low-speed network, there are fewer exploration rate adjustment epochs compared to the fast-moving network.
Likewise, Figure \ref{fig:ener_speed} presents the packet transmission energy for different networking scenarios \myblue{in each curve}. Abrupt changes in the energy consumption are corresponding to selecting a different optimal path by the algorithm. We can see that for a network with slow-moving nodes, we experience fewer sharp transitions, compared to the network with faster nodes, as expected. This shows the reasonable operation of the proposed method. It is noteworthy that we observe fewer fluctuations for the baseline method with \myblue{non-adaptive $T$, compared to the proposed method for high-speed networks.} This is not necessarily a desirable behavior since it can lead to selecting non-optimal paths by missing the newly emerged optimal links when the network topology changes drastically.
\myblue{Therefore, extremely fast switching to choose the optimal path may go against the reliability and stability of the communication. In Figure \ref{fig:stability}, we analyze the performance of the communication system to show that stability and reliability are not a concern. We use a slow network setup to see the effect of changing paths at the lowest dynamic rate. As shown, we observe that the proposed method with adaptive exploration rate selects the optimal path. In contrast, the routing protocol which uses SA with non-adaptive temperature misses the optimal links. Also, if we study the impact of switching paths on the stability and reliability of the network, we observe that in this case, the \textit{non-adaptive} approach has more variability in choosing paths than the proposed method. Therefore, for our design, changing paths will depend on the variability of the network to not miss optimal paths when the network topology changes, and not compromising the stability of the communication.}


\begin{figure}[h]
	\centering
	\includegraphics[width=0.9\columnwidth]{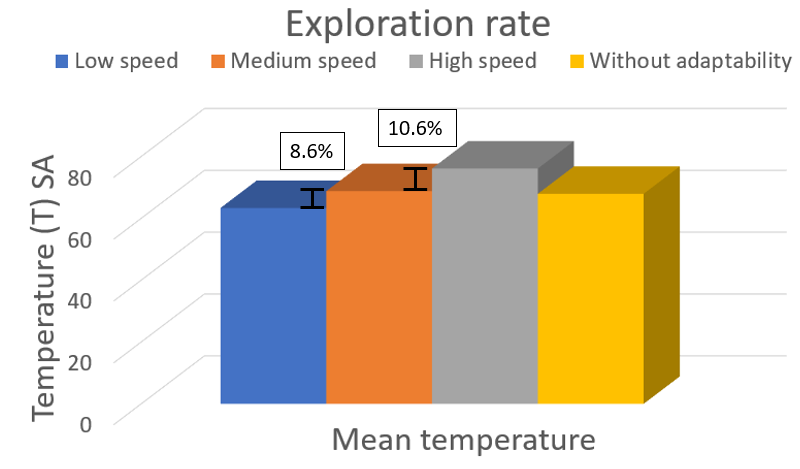}
	\caption{\small The average of the temperature parameter $T$ over time for the proposed method, as well as the baseline method with non-adaptive temperature \cite{simann_q_routing} for networks with different average speeds.}
    \label{fig:temp_graph}
\end{figure}

\myblue{In Figure \ref{fig:temp_graph}, it is seen} that the average temperature ($T$) value for the SA algorithm increases for more dynamic networks with faster mobile nodes. It implies that we explore more often for faster networks to find newly emerged optimal paths since $T$ is proportional to the exploration rate. \myblue{We observe that the average temperature varies around $8.6\%$ from low speed to medium speed networks and around $10.6\%$ from medium to high-speed networks.} Without the proposed method of controlling $T$ based on the measured Q-value change rates, we have a $T$ parameter that always declines over time and hence misses the opportunity of discovering new paths when the network topology undergoes abrupt changes. 



\myblue{Finally, we compare the performance of the proposed inferred SA method against other heuristic algorithms, such as GD, GA, and PSO, under different average node speeds. The results are presented in Table \ref{table_heuristic}. All heuristic algorithms offer a similar outcome in terms of the average transmission energy.
However, the proposed modified SA method, where the exploration rate is controlled by an adaptive temperature cooling process, achieves a much higher energy efficiency. Particularly, the results show that the proposed method performs around $20\%$ better than the conventional optimization algorithms in terms of energy consumption. The improvement is slightly higher for bigger networks.}

\begin{table}[h]
\caption{\small Comparative analysis for different heuristic algorithms in terms of the average transmission energy.}
\label{table_heuristic}
{\footnotesize
\begin{tabular}{|p{1.8cm} | p{0.7cm} | p{0.8cm} | p{0.7cm} | p{0.7cm}  |p{0.8cm} | p{0.7cm} |} 
\hline
& \multicolumn{3}{c|}{ $N$ = 10} & \multicolumn{3}{c|}{ $N$ = 20}\\
\hline
 & Slow & Medium & Fast & Slow & Medium & Fast\\
\hline
PSO / GA / GD & 83.28 & 79.67 & 79.70 & 76.98 & 78.95 & 72.64 \\
\hline
Proposed (adaptive SA) & 67.26 & 62.18 & 64.26 & 58.72 & 57.44 & 55.64 \\
\hline
\textbf{Gain} & \textbf{19.23\%} & \textbf{21.95\%} & \textbf{19.37\%} & \textbf{23.72\%} & \textbf{27.25\%} & \textbf{23.40\%} \\ 
\hline
\end{tabular}
}
\vspace{-0.15 in}
\end{table}

\section{Conclusion}   \label{sec:conclusion}

In this work, we introduced a novel fully-echoed Q-routing protocol with adaptive learning rates optimized by the Simulated Annealing algorithm based on \ar{the} inferred level of network dynamicity. 
Our method improves upon different implementations of Q-routing in terms of the convergence rate, the optimality of the end solution, and the adaptability to the network dynamicity. This gain is achieved by controlling the exploration rate of the Q-routing by regulating the temperature cooling rate of the utilized SA optimization based on the variation rate of the Q-values. 
Simulation results suggest that our algorithm achieves a reduction in the energy consumption between 7\% to 82\% and an increase of up to 264\% as for a successful packet delivery ratio, compared to other Q-routing algorithms. \ar{The choice of SA is essential since the proposed SA-based method with adaptive temperature cooling process outperforms the same routing algorithm with other heuristic optimization methods, including GA, GD, and PSO.} The proposed algorithm can solve the two key issues of UAV networks, namely the limited energy consumption and the network connectivity loss. 
\ar{Developing a more formal way of predicting per-node mobility and incorporating it into the optimization framework can be pursued as a future direction. A potential approach would be inferring the velocity of the nodes based on how fast the metrics change using a Bayesian inference method and using it to update Q-values in online Q-routing protocols.}

\vspace{-0.1 in}

\bibliographystyle{IEEEtran}
\bibliography{main}{}
\renewcommand{\baselinestretch}{0.85}

\vspace{-1.6cm}

\begin{IEEEbiography}[{\includegraphics[width=1in,height=1.25in,clip,keepaspectratio]{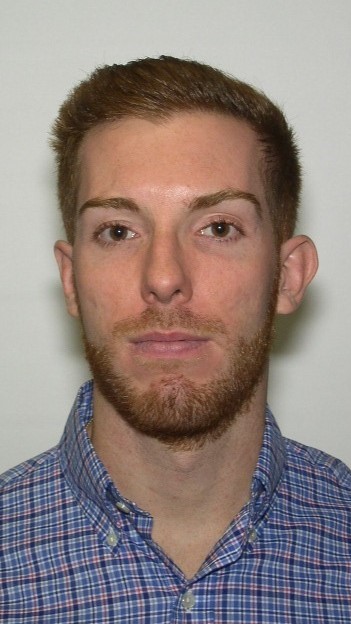}}]%
{Arnau Rovira-Sugranes}
is a PhD. candidate in Informatics and Computing in the School of Informatics, Computing and Cyber Systems (SICCS) in Northern Arizona University (NAU), Flagstaff, AZ, USA. He graduated with a Degree in Industrial Electronics and Automation Engineering from Rovira i Virgili University in 2016, coursing his senior year Electrical Engineering in Northern Arizona University as an exchange student. After graduating, he worked in a automotive seating and electrical systems company before starting his PhD. program.
His research interests include machine learning and data mining, communication and routing protocols and graph theory for flying ad-hoc wireless networks (FANETs), which produced peer-reviewed publications on predictive based solutions for Internet of Things. Also, he has served as an IEEE conference/journal paper reviewer and symposium moderator.
\end{IEEEbiography}

\vspace{-1.5cm}

\begin{IEEEbiography}[{\includegraphics[width=1in,height=1.25in,clip,keepaspectratio]{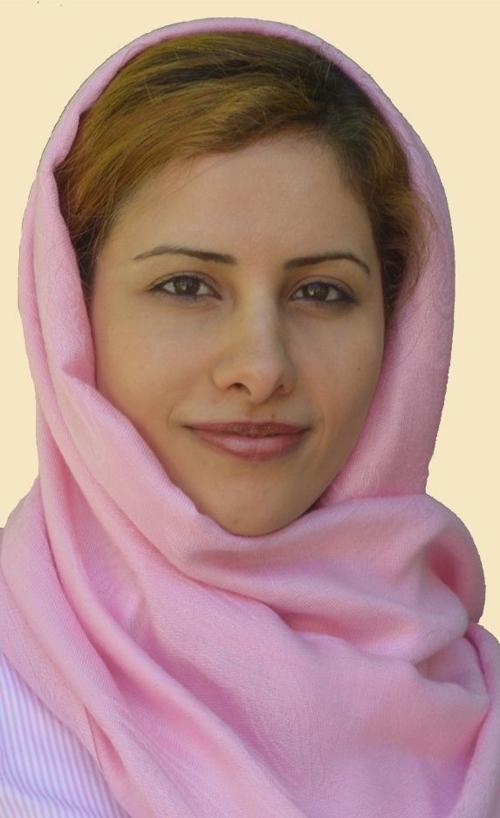}}]%
{Fatemeh Afghah}
 is an Associate Professor with the School of Informatics, Computing, and Cyber Systems, Northern Arizona University (NAU), Flagstaff, AZ, USA, where she is the Director of Wireless Networking and Information Processing (WiNIP) Laboratory. Prior to joining NAU, she was an Assistant Professor with the Electrical and Computer Engineering Department, North Carolina A$\&$T State University, Greensboro, NC, USA, from 2013 to 2015. Her research interests include wireless communication networks, decision making in multi-agent systems, radio spectrum management, and artificial intelligence in healthcare. Her research has been continually supported by NSF, AFRL, AFOSR, and ABOR.
 She is the recipient of several awards including the Air Force Office of Scientific Research Young Investigator Award in 2019, NSF CAREER Award in 2020, NAU's Most Promising New Scholar Award in 2020, and NSF CRII Award in 2017. She is the author/co-author of over 80 peer-reviewed publications and served as the organized and the TPC chair for several international IEEE workshops in the field of UAV communications, including IEEE INFOCOM Workshop on Wireless Sensor, Robot, and UAV Networks (WiSRAN’19) and IEEE WOWMOM Workshop on Wireless Networking, Planning, and Computing for UAV Swarms (SwarmNet’20).
 \end{IEEEbiography}
 
\vspace{-1.5cm}
\begin{IEEEbiography}[{\includegraphics[width=1in,height=1.25in,clip,keepaspectratio]{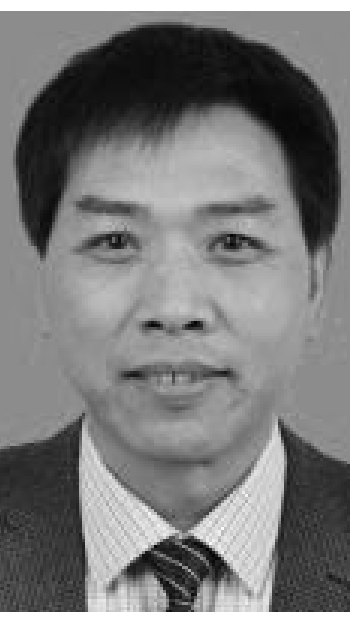}}]%
{Junsuo Qu}
 received the B.S. degree in telecommunication engineering from the Chongqing University of Posts and Telecommunications, Chongqing, China, in 1991, and the M.S.degree in communication and information systems from Xidian University, Xi’an, China, in 1998. He is currently a Full Professor with the School of Automation, Xi’an University of Posts and Telecommunications and a member of the China Institute of Communications. He is also the Director of the Xi’an Key Laboratory of Advanced Control and Intelligent Process. He is leading an IoT Research Team with the School of Automation.
\end{IEEEbiography}
\vspace{-1cm}
\vspace{-1cm}
\begin{IEEEbiography}[{\includegraphics[width=1in,height=1.25in,clip,keepaspectratio]{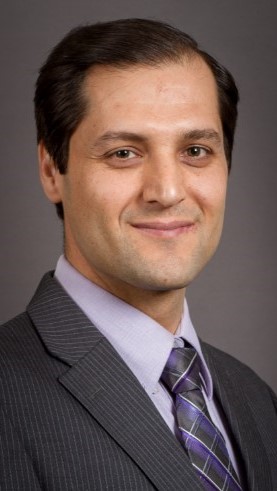}}]%
{Abolfazl Razi}
 is an assistant professor in the School of Informatics, Computing and Cyber Systems at Northern Arizona University (NAU). He received his B.S., M.S. and PhD degrees, all in Electrical Engineering, respectively from Sharif University (1994–1998), Tehran Polytechnic (1999–2001), and University of Maine (2009–2013). Prior to joining NAU, he held two postdoctoral positions in the field of machine learning and predictive modeling at Duke University (2013–2014), and Case Western Reserve University (2014–2015). He is the recipient of several competitive awards including the Best Research of MCI in 2008, Best Graduate Research Assistant of the Year Award from the College of Engineering, University of Maine in 2011, and the Best Paper Award from the IEEE/CANEUS Fly By Wireless Workshop in 2011. His current research interests include smart connected communities, biomedical signal processing, wireless networking, Internet of things, and predictive modeling.
\end{IEEEbiography}

\vspace{-1cm}

\end{document}